\documentclass[acmsmall]{acmart}
\usepackage{import}
\usepackage{enumitem}
\usepackage{multirow}
\AtBeginDocument{%
  \providecommand\BibTeX{{%
    \normalfont B\kern-0.5em{\scshape i\kern-0.25em b}\kern-0.8em\TeX}}}
\acmSubmissionID{7932}
\begin{document}
\title{The Effect of Sociocultural Variables on Sarcasm Communication Online}
\author{Silviu Vlad Oprea}
\affiliation{%
  \institution{School of Informatics, The University of Edinburgh}
  \city{Edinburgh}
  \country{UK}}
\email{silviu.oprea@ed.ac.uk}
\author{Walid Magdy}
\affiliation{%
  \institution{School of Informatics, The University of Edinburgh}
  \city{Edinburgh}
  \country{UK}}
\email{wmagdy@inf.ed.ac.uk}
\begin{abstract}
Online social networks (OSN) play an essential role for connecting people and allowing them to communicate online. OSN users share their thoughts, moments, and news with their network. The messages they share online can include sarcastic posts, where the intended meaning expressed by the written text is different from the literal one. This could result in miscommunication.
Previous research in psycholinguistics has studied the sociocultural factors the might lead to sarcasm misunderstanding between speakers and listeners. However, there is a lack of such studies in the context of OSN.  In this paper we fill this gap by performing a quantitative analysis on the influence of sociocultural variables, including gender, age, country, and English language nativeness, on the effectiveness of sarcastic communication online.
We collect examples of sarcastic tweets directly from the authors who posted them. Further, we ask third-party annotators of different sociocultural backgrounds to label these tweets for sarcasm.
Our analysis indicates that age, English language nativeness, and country are significantly influential and should be considered in the design of future social analysis tools that either study
sarcasm directly, or look at related phenomena where sarcasm may have an influence. We also make observations about the social ecology surrounding sarcastic exchanges on OSNs. We conclude by suggesting ways in which our findings can be included in future work.
\end{abstract}
\maketitle
\textbf{This is a preprint of an article accepted for publication by CSCW 2020.}
\section{Introduction}
\label{section:introduction}
Sarcasm is a form of verbal irony that occurs when there is some discrepancy between the literal and intended meanings of an utterance. Using the discrepancy, the speaker expresses dissociation towards a previous proposition, in the form of surface contempt or derogation~\citep{WILSON20061722}, although the communicative purpose may also be to praise or show appreciation~\citep{role-pexman-2004}.

Sarcasm is omnipresent on online social networks (OSN) and can be highly disruptive of systems that harness this data to detect behavioural and social dynamics signals~\citep{sarcasm-in-sentiment-analysis}, often the subject of research in the CSCW community. These signals drive crucial marketing, administration, and investment decisions~\citep{sentiment-analysis-in-business}. As such, computational systems that are able to automatically detect sarcasm in natural language utterances~\citep{joshi:2017:sarcasm-survey} can find wide usage, both within the community and beyond.

Most work on computational sarcasm detection focuses on extracting lexical and pragmatic cues available in the utterance being classified~\citep{campbell2012,riloff2013,joshi2016,tay2018}. However, previous research in psycholinguistics points out significant influences that speaker and audience sociocultural backgrounds can have on whether the message intended by the speaker is accurately perceived by the audience, i.e. on whether the communication is effective. Speaker traits, along with their familiarity with the audience, can influence both the speaker's tendency to use sarcasm, as well as the way they formulate the sarcastic utterance~\citep{gibbs-2000,rockwell-context-2001,ivanko-2004}. Similarly, listener traits can determine their predisposition to interpret the utterance as sarcastic~\citep{jorgensen-gender-1996} and awareness of speaker traits can cue the sarcastic intent of the speaker to the listener~\citep{char-occupation-1997,char-occupation-2000,char-gender-katz-2001,char-age-harris-2001,char-occupation-stereotypes-2002}. 
These studies suggest that sociocultural backgrounds of the interlocutors should be considered in sarcasm detection pipelines. However, there is a shortage of quantitative empirical evidence---especially in the context of OSN---regarding the specific sociocultural variables that should be considered, and the degree of influence that each variable has. This makes it unclear how sociocultural backgrounds should be encoded in computational models trained on web-scale datasets.

In this work we provide a quantitative investigation into the sociocultural dimensions of sarcasm, using OSN data as a lens. We formulate the following research questions:
\begin{enumerate}
    \item Is the effectiveness of sarcastic communication influenced by whether the interlocutors have similar sociocultural backgrounds?
    \item If so, which sociocultural variables have the most influence?
    \item Earlier research (e.g. \citep{coolness}) suggests that sociocultural variables may only have an influence on the effectiveness of sarcastic communication when other contextual cues to speaker intent are missing. Does this also apply to OSN online sarcastic communication?
\end{enumerate}

To answer our questions, we first choose a set of sociocultural variables to investigate by looking at both (a) psycholinguistic studies of sarcastic communication, and (b) linguistic theories of sarcasm. We then use a crowdsourcing platform to collect a dataset of tweets labelled for sarcasm by their authors. We refer to these labels as \emph{intended sarcasm} labels, since they reflect the authorial sarcastic intention.
Next, we form several treatment groups of annotators, each group containing representatives of a specific sociocultural background. Finally, in two experimental settings, we task each group with labelling the tweets in our dataset as either sarcastic or not sarcastic, referring to any such resulting label as a \emph{perceived sarcasm} label. In the first setting, the annotators are presented with the text of the tweets they are asked to label. In the second setting, they are not shown the text, but are provided with the link to the tweet, and are asked to also consider surrounding tweets and user profile information when deciding on the label.

In each experimental setting, we compare intended sarcasm labels with perceived labels from across treatment groups, to determine the sociocultural variables that have the most significant influence on whether sarcasm is perceived as intended.
We find age, English language nativeness, and gender, to have a significant influence on sarcasm perception. While the presence of contextual information alleviates the influence of English language nativeness and gender, age remains significantly influential.

Our results suggest that the inclusion of sociocultural variables in computational models of sarcasm can increase modelling performance.
Through this work, we hope to motivate a new direction of research that gives significant consideration to the influence that the sociocultural backgrounds of the interlocutors can have on sarcastic communication online.

We believe our results, and the direction of research they suggest, can have significant impact on the design of future social analysis tools deployed in settings where sarcasm can be disruptive, making our work relevant to the CSCW community. The obvious application is in analysing how people use and understand sarcasm and related non-literal linguistic phenomena across cultures. However, there are also implications for the design of downstream tools that analyse sentiment, opinions, trends, or classify hate-speech, to name just few areas where sarcasm can lead to erroneous results.

\section{Socio-Cultural Variables}
\label{section:socio-cultural-variables}
In this section we look at both (a) psycholinguistic studies of sarcasm usage and comprehension, and (b) linguistic theories of sarcasm. Our purpose is to select, for later quantitative investigation, a set of sociocultural variables that could influence the effectiveness of sarcastic communication.
We begin by defining the terminology that we use in this paper.
\subsection{Nomenclature}
Our work focuses on analysing sarcasm sharing and understanding on OSNs. While our work is directed to the CSCW community, it touches on previous research related to sarcasm across multiple domains, including linguistics, psycholinguistics, and natural language processing (NLP).
To avoid ambiguity, we provide basic definitions for the terms used later.

An \emph{interlocutor} is a participant in a conversation. We assume a conversational setting where at any specific moment we can differentiate between a \emph{speaker} and an \emph{audience}. A \emph{listener} is a member of the audience who pays attention to, and may subsequently engage with, the speaker. We say there is an \emph{effective} communication between the speaker and the listener when the communicative goal is reached. That is, the message is perceived by the listener as intended by the speaker.

Each interlocutor is characterised by a \emph{sociocultural background}, which is a set of sociocultural \emph{traits}. Such a trait is an instance of a \emph{sociocultural variable} that determines a coarse partition over the space of potential interlocutors. That is, it does not identify any individual interlocutor, but a set of them. We say such a variable \emph{influences} a communication unit if the effectiveness of the unit would change if one or more interlocutors would possess a different value for that variable than they do.

Since we have defined a sociocultural background as a set, we further employ set-theoretic terminology to compare and contrast backgrounds. As such, the similarity between two backgrounds is quantified as the cardinality of their intersection. As special cases, we say two backgrounds are the same if they fully overlap, and are disjoint if their intersection is the empty set.
%
\subsection{Psycholinguistic Studies of Sarcastic Communication}
\label{section:socio-cultural-variables:psycholinguistic}
We begin by considering sociocultural variables that have been suggested to influence sarcasm usage and comprehension in previous psycholinguistic studies.
%
\subsubsection{Gender}
The influence of gender on the tendency to use sarcasm has been studied widely in psycholinguistics. \citet{gibbs-2000} notices males to be more likely to use sarcasm in conversations with friends than females.
\citet{char-gender-katz-2001} investigate whether gender of the speaker could cue their tendency to use sarcasm to the listeners. They find that males are perceived to be more sarcastic than females. Building upon their work, \citet{char-gender-taylor-2016} notice no correlation between gender and tendency to \emph{use} sarcasm. Rather, they notice a preference for perceiving male behaviour as sarcastic. \citet{jorgensen-gender-1996} look at gender differences in emotional reactions to sarcasm. They notice males to be more likely than females to perceive humor in sarcasm, and females to be more likely to be offended or angered by sarcasm.
%
\subsubsection{Age}
Developmental literature suggests children begin to use personality traits to infer non-obvious meanings as young as age 4~\citep{char-age-heyman-1999}. \citet{char-age-harris-2001} explore children's abilities to use the speaker traits as cues to sarcastic intent. They look at whether consistent personality trait information, such as being told a sarcastic criticism was made by a mean speaker, would enable the detection of sarcastic intent. 
They notice that younger children rely more heavily on trait information, while older children have a stronger understanding of the phenomenon of sarcasm and a more complex way of integrating speaker traits into the discerning process. Their work supports the hypothesis that age can be a determining factor in the effectiveness of sarcastic communication when the listeners are children. We are unaware of similar studies on adults. We investigate whether the trend holds for adults of different ages in Section~\ref{section:results-and-analysis}.
%
\subsubsection{Country}
In a quantitative study, \citet{joshi-etal-2016-cultural} present a dataset of tweets, initially labelled for sarcasm by American annotators, to also be labelled by Indian annotators. They find higher disagreement between annotators of different nationalities, than between annotators of the same nationality. They consider cultural differences between India and the United States to be the cause of this disagreement. There is a subtle aspect that makes their work very different from ours. Mainly, they compare the labelling disagreement between different listeners of sarcasm, not the ability of listeners to effectively perceive sarcasm as intended by the speakers of the tweets in their dataset. Our purpose, however, is to assess precisely this ability. In fact, their dataset lacks annotations for the sarcastic intention of the speakers of the tweets altogether. Nevertheless, we are compelled by their results and wish to include country as a variable of investigation. In our work, we choose to investigate two countries: the United Kingdom (UK) and the United States (US). There are two reasons for choosing these. First, research~\citep{us-vs-uk} suggests that the two countries, despite sharing the same language, adopt different methodologies for using and interpreting linguistic constructs. This may extend to sarcasm usage. Second, there is a pragmatic aspect to our choice, motivated by the prevalence of US and UK workers on an online crowdsourcing platform that we use later to collect and label data.
%
%
\subsection{Linguistic Theories of Sarcasm}
\label{section:socio-cultural-variables:linguistic}
In this section we consider whether linguistic theories of sarcasm can account for the influence of sociocultural variables on the effectiveness of sarcastic communication. Linguistic theories do not make explicit predictions about the role of such variables. However, as pointed out by~\citet{coolness}, we may attempt to derive predictions from the basic assumptions of each theory.
%
\subsubsection{Grice}
\label{section:sociocultural-variables-grice}
One of the first formal accounts for sarcasm is provided by Grice, who views it as a flouting of the first maxim of Quality~\citep{grice-1975}. Here, a flouting is a blatant violation that gives rise to a conversational implicature. In other words, in Grice's view, the speaker of a sarcastic utterance does not figuratively mean, but conversationally implicate the opposite of what they say. That is, ``what a great movie'' conversationally implicates that the movie was bad. 
A main limitation of the Gricean view is that the flouting is neither necessary nor sufficient for sarcasm to occur. To see that it is not necessary we can consider sarcastic understatements, such as saying "This was not the best movie ever" to mean the movies was bad. To see that it is not sufficient, we simply note that not every utterance that is literally false is sarcastic. An example in this direction are metaphors.
Despite this disadvantage, (discussed in more detail by~\citet{sperber-1981}), there are examples of sarcasm that this view does explain, such as those offered by~\citet{grice-1975}. Let us return to our purpose of finding sociocultural variables that may influence sarcastic communication. While the Gricean view does not make direct predictions for such variables, it does imply that the listener of sarcasm needs to be able to implicate the meaning intended by the speaker. In this direction, research~\citep{bouton-1988,bouton-1992} find that non-native speakers of English tend to differ in the meaning they attribute to conversational implicatures compared to native speakers, in a context where English language proficiency is controlled. Based on this research, English language nativeness of the interlocutors is a socio-cultural variable that we expect to have an influence the effectiveness of sarcastic communication.
We invite the interested reader to consult~\citet{grice-1975} and \citet{WILSON20061722} for more details on the Gricean view of sarcasm.
%
\subsubsection{Sarcasm as an Indirect Speech Act}
\label{section:sociocultural-variables-indirect-speech-acts}
Sarcasm can also be viewed through the lens of the speech act theory~\citep{austin:1962:speech-acts}. Speech acts are acts performed by speaking, i.e. acts performed by the propositions that our utterances express. Such acts include requesting, asking, promising, and blaming. If the literal and non-literal meanings of an utterance are not performing the same acts, then the non-literal meaning is referred to as an indirect speech acts~\citep{searle:1975:indirect-speech-acts}. Searle formulates a set of felicity conditions that effective speech acts should meet~\citep{searle:1969:speech-acts}. That is, a set of rules that must be met for a speech act to achieve its purpose.
Based on Searle's work, \citet{amante:1981:ironic-speech-act} sees sarcasm as blatant failure to satisfy one or more of these conditions. That is, sarcastic language is deceptive, but superficially so, the speaker intending to expose their infelicitous act to the listener. Consider the utterance ``Brilliant job!'' addressed by Alice to Bob after looking in the oven to discover an overbaked cake forgotten about by Bob. Alice violates the \emph{preparatory} felicity conditions, which ask the speaker to have both evidence for the truth of their proposition $P$, and knowledge that the listener is not aware of $P$.
Indeed, Alice has no reason to believe her proposition $P$ (i.e. that it was brilliant that the cake was overbaked). In fact, she has evidence for the negation of $P$. She also knows that Bob considers the negation of $P$ to be the case, rendering the literal sense of her utterance redundant. Alice also violates the \emph{sincerity} rule, which asks her to believe $P$ to be true. Her statement is, thus, infelicitous, and she offers cues to her intentions, e.g. by using ``brilliant'', word that is often part of hyperbole.
The result is the construction of a latent meaning that stands in antithesis to the literal one, i.e. a critique about Bob's forgetfulness.
This view of sarcasm suffers from limitations similar to those of the Gricean view. First, violation of felicity conditions is not necessary for sarcasm to occur, as argued by~\citet{colston:2000:conditions} and \citet{utsumi-2000}. In this direction consider again sarcastic understatements. Violation is also not sufficient, as it does not provide grounds to discriminate between sarcasm and other indirect speech acts, such as metaphors\footnote{See \citet{jakub:2012:searle-metaphor} for an analysis of Searle's view of metaphors under the speech act theory.}.

However, for the instances of sarcasm that it does explain, we return to our goal of identifying sociocultural factors that may influence the effectiveness of such instances. In this direction, \citet{searle:1975:indirect-speech-acts} notes that the apparatus needed to explain indirect speech acts includes knowledge of the mutually shared background information of the interlocutors. This includes shared social norms and expectations, all of which can contribute to the successful communication of sarcasm, because such information is often often evoked and negated to create sarcasm~\citep{amante:1981:ironic-speech-act}. In our example above, Alice considers it a norm that cakes should not be served overcooked. For her sarcasm to be understood, it is essential for Bob to share this view. As such, we expect sarcastic exchanges to be more effective when the interlocutors have the same sociocultural background, compared to when they do not, a hypothesis that we test quantitatively in Section~\ref{section:results-and-analysis}.

\citet{amante:1981:ironic-speech-act} further notes that, while the social norm or expectation evoked could co-occur with its negation in the same utterance, two two could also be separated by large distances in the exchange. When the interlocutors have sociocultural backgrounds that are so different that the evoked norm or expectation is not shared among them, one could expect contextual information beyond the sarcastic utterance to be particularly useful for indicating the sarcastic intention of the speaker to the listener. This is also conjectured, but not validated, by~\citet{coolness}. We provide an analysis in this direction in the context of OSNs.
%
\subsubsection{Echoic Theories}
Consider the sarcastic utterance ``what a great movie'' spoken after a movie the speaker thought was bad. \citet{sperber-1981,sperber-relevance-1986} offer a different account of sarcasm. They argue that the purpose of the sarcastic utterance cannot be to convey the belief that the movie was bad, since the belief can only be understood from the utterance if we know the utterance is sarcastic. However, we can only know it is sarcastic if we know the speaker's belief in advance. This makes the utterance completely uninformative if the purpose is to convey the speaker's belief about the movie.
In the view of \citet{sperber-1981}, the speaker is trying to convey a belief not about the movie, but about the utterance ``what a great movie'' itself. The utterance is an echoic mention of the speaker's initial expectation to see a good movie.
This \emph{echoic mention theory} of sarcasm explains why sarcastic utterances are made and why the meaning they implicate can be incongruous to the literal meaning. Further, is does not require a mechanism for pointing out an implicature.
However, it does not differentiate between sarcastic and non-sarcastic echoic mentions.
\citet{kreuz-1989} address this limitation by introducing the \emph{echoic reminder theory} of sarcasm which adds the constraint that the echoic mention should always remind the listener of a violated social norm or a failed expectation.
Echoic theories allow for listener traits to play a role. As \citet{coolness} argue, some listeners, perhaps more cynical, might attend more to failed expectations, therefore expecting to see sarcasm more often in conversations than others.
Off course, this might lead to errors in communication, if the interlocutors have different ideas of what constitutes the failed state of an expectation. In this direction, we expect listeners that have similar sociocultural backgrounds with the speaker to perceive sarcasm as intended by the speaker more accurately than listeners with backgrounds that are different from that of the speaker. We test this in our experiments in Section~\ref{section:results-and-analysis}.
%
\subsubsection{Pretense Theories and the Common Ground}
\citet{clark-1984} introduce the \emph{pretense theory} of sarcasm which claims that a sarcastic speaker pretends to be an injudicious person speaking to an imaginary uninitiated audience who would interpret their utterance literally. This way the speaker expresses a negative attitude towards the pretended injudicious person, the imaginary audience, and the situation portrayed through their acting.
The actual listener is expected to discover the pretense and this way understand the sarcasm.
A variant of the pretense theory is considering sarcasm a pretense that the interlocutors jointly perform. That is, they both act like performing a serious communication act in an imaginary situation. Their joint pretense that this situation is taking place is what generates sarcasm. An implication that constitutes a main limitation of this approach, as pointed out by~\citet{utsumi-2000}, is that the listener needs to share the sarcastic intention with the speaker beforehand, so that they (the listener) can engage in the joint pretense. Another limitation, shared with the original pretense theory, is the failure to distinguish between sarcastic and non-sarcastic pretense. An example of the latter is parody.

For instances of sarcasm that the pretense theories do explain, socio-cultural traits may play a role in so far as these traits are consistent with this type of behaviour. First, as suggested by~\citet{coolness}, if the speaker exhibits traits that are consistent with insincerity and injudiciousness, then the listener might be more inclined to expect sarcasm from the speaker.
Second, as~\citet{coolness} point out further, if the listener shares these traits, they may be more able to detect the speaker's pretense. As such, similar to last section, this seems to suggest that the most efficient setting for sarcastic communication is when the the speaker and listener have similar sociocultural backgrounds.

In the category of pretense theories there is also the recent work of \citet{gordon:2019:cool}. They suggest viewing sarcasm as a form of linguistic countersignaling~\citep{feltovish:2002:countersignalling}: a communicative act where the interlocutors engage in a joint pretense about the state of the world, or the perspective that they hold, with the purpose of communicating about the common ground~\citep{grounding-clark-1984,grounding-clark-1985,fugelli:2013:intersubjectivity}. The effectiveness of the exchange can, thus, be quantified by considering amount of the shared knowledge that they have~\citep{grounding-clark-1985}. This leads us to the same expectation of a more effective exchange between interlocutors of similar sociocultural backgrounds, under the assumption that such backgrounds determine social partitions. We investigate this in Section~\ref{section:results-and-analysis}.
What sets out work apart from previous attempts at analysing the effect of common ground on the effectiveness of sarcastic communication is that, to our knowledge, previous analyses, e.g.  \citep{grounding-clark-1984,grounding-clark-1989-traits}, (a) do not consider the sociocultural variables that we do; (b) conduct exclusively qualitative analyses; and (c) do not analyse OSN exchanges. The qualitative investigation of~\citet{grounding-osn-qualitative-2017} does analyse the extent to which interlocutor common ground affects the effectiveness of OSN communication, but their research does not consider sarcasm, nor our variables of interest.
\subsubsection{Implicit Display Theory}
\label{section:socio-cultural-variables:implcit-display-theory}
\citet{utsumi-2000} argues that none of the theories discussed so far provides a complete account of sarcasm, in that the conditions they presuppose are neither necessary, nor sufficient, for sarcasm to occur.
As an alternative, Utsumi suggests the \emph{implicit display theory} of sarcasm, which focuses on explaining how sarcasm is distinguished from non-sarcasm. The theory defines three criteria that are necessary for sarcasm to occur. First, the communicative context should include elements that constitute an \emph{ironic environment}, that is, a situation that motivates the use of sarcasm. Second, the sarcastic utterance should \emph{implicitly display} the ironic environment. Third, the listener should assign the utterance a degree of sarcasm that is proportional to the degree to which the utterance achieves implicit display of the ironic environment. That is, sarcasm is a prototype-based category. An utterance is more or less sarcastic based on how little or much it deviates from prototypical sarcasm. The prototype is that instance of sarcasm which implicitly displays the ironic environment most accurately.
Utsumi provides three conditions that are necessary for a situation in which an utterance is given to be surrounded by an ironic environment. First, the speaker of the utterance has an expectation $E$ at time $t_0$. Second, $E$ fails, i.e. is incongruous with reality, at time $t_1$. Third, the speaker has a negative emotional attitude towards the incongruity between what they expected and what actually happened.
\citet{coolness} points out several ways in which sociocultural traits could influence sarcasm usage and perception under this theory. First, traits of the speaker could cue what expectations they usually have, as well as how they express negative attitudes. Second, there might be traits associated with prototypical sarcasm, such as being likely to react negatively and to think critically. On one hand, if the listener is aware that the speaker possesses these traits, the listener might perceive the speaker's utterance closer to the prototype. On the other hand, if the listener possesses these traits, that could make them more likely to judge the utterance as closer to the prototype. As in the previous sections, the setting that assures the maximum amount of accurate information transfer could be when the speaker and the listener have similar sociocultural backgrounds.
%
%
\section{Sarcasm and Trolling}
\citet{trolling-2014} suggest that sarcasm is a type of trolling. From here, one could further consult the work of~\citet{trolling-2016}, and that of~\citet{trolling-2017}, who show that two of the variables we have selected for investigation, age and gender, are associated with trolling. This would render our investigation in that direction redundant. However, based on previous linguistic and psycholinguistic studies of sarcasm, we find it problematic to consider sarcasm a type of trolling. To show why this is the case, we show that the intention to troll is not necessary for sarcasm to occur.

From a formal linguistic perspective, the argument is straightforward. Note that none of the linguistic theories discussed in Section~\ref{section:socio-cultural-variables:linguistic} presuppose an intention to troll as a prerequisite for constructing sarcasm. Grice's theory only postulates the violation of a maxim and nothing about how that violation is achieved. Echoic theories have no claim over the manner in which dissociation from a previous proposition is achieved. While the Implicit Display Theory requires an expression of a negative attitude, it does not require that attitude to be trolling. In fact, it does not require that the expression should have an addressee at all. Indeed, it could well be directed at an object, or could be self-reflexive.

The fact that trolling is not necessary for sarcasm to occur is even more apparent if we look at psycholinguistic studies into the role of sarcasm in communication. Of particular relevance are the works of~\citet{role-jorgensen-1996} and~\citet{role-pexman-2004}, who argue that one of the reasons a speaker might choose to use sarcasm is to demonstrate and enhance relationship closeness with the listener, as a linguistic code between friends, or to show affection or appreciation.

As such, the assumption of~\citet{trolling-2014} that sarcasm is a type of trolling cannot hold. This gives us reasons to question whether the claims of \citet{trolling-2016} and \citet{trolling-2017} about the roles of age and gender in trolling apply to sarcasm. This further motivates the current work.
%
%
\section{Data Collection \& Analysis Methodology}
\label{section:data-collection}
Our purpose is to quantitatively investigate the influence of the sociocultural variables introduced in Section~\ref{section:socio-cultural-variables} on the effectiveness of sarcastic communication on OSN. We use Twitter data annotated for intended and perceived sarcasm for our investigation. In this section, we describe the process of collecting our tweets along with the corresponding labels.
\subsection{Collecting the Tweets with Intended Sarcasm Labels}
\label{section:data-collection:intended-sarcasm}
We designed an online survey and published it on the Prolific Academic\footnote{https://prolific.ac} crowdsourcing platform, where we asked Twitter users to provide links to one sarcastic and three non-sarcastic tweets that they had posted in the past, on their own timeline, or as replies to other tweets. The labels are thus implicitly specified by the authors themselves, representing their sarcastic intention (intended sarcasm).

We implemented quality control steps to prevent spurious entries and decrease the chance that the users might misjudge the sarcastic nature of their previous tweets under experimental bias. As such, we made sure all four tweets submitted by a participant in a survey response were in English and were posted at least 48 hours before the survey submission, to avoid participants posting tweets on the spot. We also checked that all tweets came from the same accounts that should be unverified, with less than 30k followers, to avoid participants submitting tweets from famous accounts.
In addition, we asked users to explain what made their sarcastic tweet sarcastic and to provide a rephrase of their tweet would convey the same message non-sarcastically. Survey responses done in less than 3 minutes were disregarded.
The 3 minute cut off was chosen after many test iterations of data collection followed by manual inspection. We manually noted that survey responses tended to be of higher quality if the participants took at least 3 minutes filling them in. To assess the quality, we looked at the explanation and rephrases of sarcastic tweets.
In a further quality control step, we provided the tweets along with explanations and rephrases to a linguistic expert, asking them to filter out from our dataset all tweets that, despite being declared as sarcastic by their authors, were obvious noise. That is, they were most likely provided just to receive payment for filling in our form.
The contributors have agreed to have the IDs of the tweets they provided, as well as their intended sarcasm labels, made public as part of open science, and that we may collect public information from their profile. The participant information sheet shown to the contributors has been designed in conjunction with the ethics committee at our institution after receiving the required IRB approval for the study.
Table~\ref{table:examples} shows a sample of sarcastic tweets from our dataset.
\begin{table}[t]
    \centering
    \scriptsize
    \begin{tabular}{@{}p{4cm}p{5.6cm}ccc@{}}
        \toprule
            \textbf{tweet text} & \textbf{explanation} & \textbf{country} & \textbf{gender}\\\midrule
            I've spent over an hour trying to decide which font to use for a poster, so I think it's fair to say I'm making some pretty valuable contributions to Science today. & This tweet is sarcastic because spending an hour deciding on a font was procrastination from creating a poster for a scientific conference, and I did not make a single valuable contribution to science (which is my job) that day. & UK & female\\\midrule
            Jurassic World 2 trailer at 2am? Never have I been so grateful to have sleep problems & I hate that I have sleep problems and I would have been much happier being asleep than awake & UK & female\\\midrule
            Glad the president to be is watching snl instead of you know learning about  how to be a president & i am not glad trump cares about tv more than real issues. & US & male \\\midrule
            Guess they are not rich enough to get their precious cars in a garage. & They are rich enough to put the spikes on trees to keep birds away so they don't shit on their cars, if they can't build a house or build a garage then they are def. not rich enough. They rather hurt nature in order to keep their property "clean" than come up with an idea to fix their own problem without hurting other things. & US & male \\
        \bottomrule
    \end{tabular}
    \caption{Examples of sarcastic tweets from our datasets, labelled for sarcasm by their authors, as discussed in Section~\ref{section:data-collection:intended-sarcasm}. We also the explanations that authors gave as to what made their tweets sarcastic (explanation) and the demographic information collected.}
    \label{table:examples}
\end{table}

Prolific Academic allows targeting workers of specific sociocultural traits. Our variables of interest, discussed in Section~\ref{section:socio-cultural-variables}, are among those available. We chose to investigate the top sociocultural backgrounds in terms of the size of the partition they determine over the space of workers. As such, we targeted female workers from the United Kingdom, between 25- and 34-years-old (background referred to as F\_25-34\_UK); and male workers from the United States, of the same age (M\_25-34\_US).
We collected a total of 30 responses for each background, making a total of 240 tweets for both backgrounds, with a proportion of 1:4 of sarcastic to non-sarcastic tweets.
\subsection{Collecting Perceived Sarcasm Labels}
\label{section:data-collection:collecting-perceived-sarcasm-labels}
We now describe how we collected perceived sarcasm labels for the tweets from the previous section. Our plan was to compare intended and perceived labels, as a way of tackling our research questions. We collected perceived labels from different treatment groups, and in different settings, depending on the question addressed.

\subsubsection{First Research Question}
\label{section:data-collection:perceived-sarcasm:first-research-question}
The first question asks if the effectiveness of sarcastic communication between a speaker (Twitter user who posted one of our sarcastic tweets) and a listener (annotator providing a perceived sarcasm label for that tweet) is influenced by whether they have similar sociocultural backgrounds.
To investigate this, we published a further online survey on Prolific Academic. The survey showed the texts of several tweets and asked listeners (survey participants) to label each tweet as either sarcastic or non-sarcastic. For each tweet, we collected such labels from two treatment groups, with 3 separate labels per tweet from each group, to alleviate labelling noise. The first group consisted of listeners who had the same sociocultural background as the speaker of the tweet. The second group contained listeners of backgrounds disjoint to the background of the speaker. That is:
\begin{itemize}
    \item If the tweet came from a female speaker from the United Kingdom who is between 25- and 34-years-old (F\_25-34\_UK), the first group contained listeners of the same background (F\_25-34\_UK), while the second group contained male listeners from the United States who are over 45-years-old (M\_>45\_US);
    \item Similarly, if the tweet came from a male speaker from the United States who is between 25- and 34-years-old  (M\_25-34\_US), the first group contained listeners of the same background (M\_25-34\_US), while the second group contained female listeners from the United Kingdom who are over 45-years-old (F\_>45\_UK).
\end{itemize}
We have chosen these two age groups considering Erikson's stages of psychosocial development~\citep{erikson:1994:identity}. He defines the early adulthood stage to be between 20- and 30-years-old; and the middle and late adulthood stages over 40-years-old. We decided to introduce an offset at both ends of the age intervals, to ensure a stronger separation.
As such, our first age group is between 25- and 34-years-old, belonging to the early adulthood stage; and the second group is over 45-years-old, belonging to the middle or to the late adulthood stages.
For brevity, we introduce the following condensed notation. In a sentence, the specific background of the speaker could be irrelevant for the point that the sentence makes, i.e. it can be any of the two speaker backgrounds that we consider (F\_25-34\_UK or M\_25-34\_US). In such a scenario, we use the notation \emph{list=speak} to refer to the treatment group that contains listeners with the same background as the speaker, and \emph{list$\ne$speak} to refer to the group with listeners of opposing backgrounds. For instance, if speaker background is F\_25-34\_UK, \emph{list=speak} denotes the F\_25-34\_UK group of listeners, and \emph{list$\ne$speak} denotes the M\_>45\_US group. The treatment group notation is summarised in Table~\ref{table:treatment-group-notation}, and the condensed notation in Table~\ref{table:condensed-notation}.

We compared intended labels with \emph{list=speak} and \emph{list$\ne$speak} perceived labels across our dataset, to see which of the two groups was best at capturing sarcasm as intended by the speakers. This allows us to make a statement about our first research question, as discussed in Section~\ref{section:results-and-analysis:rq1}.
\subsubsection{Second Research Question}
\label{section:data-collection:perceived-sarcasm:second-research-question}
The second question asks which sociocultural variables have the most influence on the effectiveness of online sarcastic communication.
To investigate this, we used the same survey as we did when collecting perceived sarcasm labels for tacking the first research question. Here, we collected such labels from four treatment groups, with 3 separate labels per tweet from each group. Each group corresponds to one of the four sociocultural variables we chose for investigation: age, gender, country, and English language nativeness. In each group, the listeners have the same background as the speaker, except for flipping the corresponding variable of study. That is:
\begin{itemize}
    \item If the tweet came from a F\_25-34\_UK speaker, the first group, used for studying the influence of age, contained female listeners from the United Kingdom who were over 45-years-old (F\_>45\_UK). The second group, studying the influence of gender, contained M\_25-34\_UK listeners. The third one, studying the influence of country, contained F\_25-34\_US listeners. Finally, the forth one, studying the influence of English language nativeness, contained female listeners, between 25- and 34-years old, whose first language was not English, but declared to be fluent in English. We denote them as F\_25-34\_!native;
    \item Similarly, if the tweet came from a M\_25-34\_US speaker, the four groups were M\_>40\_US (age flipped), F\_25-34\_UK (gender flipped), M\_25-34\_UK (country flipped), and M\_25-34\_!native (non-native, but fluent, speakers of English).
\end{itemize}
Similar to the previous section, we introduce a condensed notation. In a sentence where the specific background of the speaker is not specified, we use \emph{list=speak-[variable]} to refer to the treatment group that contains listeners that have the same background as the speaker, except for the specific \emph{variable} being flipped. For instance, if speaker background is F\_25-34\_UK, \emph{list=speak-age} denotes the F\_>45\_UK group of of listeners, \emph{list=speak-gender} denotes the M\_25-34\_UK group, \emph{list=speak-country} denotes the F\_25-34\_US group, and \emph{list=speak-native} denotes the F\_25-34\_!native group. Notational conventions are summarised in Table~\ref{table:treatment-group-notation} (treatment group notation) and Table~\ref{table:condensed-notation} (condensed notation).
\begin{table}[t]
\setlength{\tabcolsep}{2pt}
    \centering
    \footnotesize
    \begin{tabular}{p{2.5cm}l}
        \toprule
            \textbf{Group notation} & \textbf{Description}\\\midrule
            F\_25-34\_UK & Females between 25- and 34-years-old from the United Kingdom\\
            F\_25-34\_US & Females between 25- and 34-years-old from the United States\\
            F\_>45\_UK & Females over 45-years-old from the United Kingdom\\
            F\_25-34\_!native & Females between 25- and 34-years-old, fluent, but non-native, speakers of English\\
            M\_25-45\_US & Males between 25- and 34-years-old from the United States\\
            M\_>45\_US & Males over 45-years-old from the United States\\ 
            M\_25-34\_UK & Males between 25- and 34-years-old from the United Kingdom\\
            M\_25-34\_!native & Males between 25- and 34-years-old, fluent, but non-native, speakers of English\\
        \bottomrule
    \end{tabular}
    \caption{Summary of the notation used to denote listener treatment groups, as discussed in Section~\ref{section:data-collection:collecting-perceived-sarcasm-labels}.}
    \label{table:treatment-group-notation}
\end{table}
\begin{table}[t]
\setlength{\tabcolsep}{2pt}
    \centering
    \footnotesize
    \begin{tabular}{l|l|l|l|l|l|l}
        \toprule
            \textbf{Speaker} & list=speak & list$\ne$speak & list=speak-age & list=speak-gender & list=speak-country & list=speak-native\\\hline
            \textbf{F\_25-34\_UK} & F\_25-34\_UK & M\_>45\_US & F\_>45\_UK & M\_25-34\_UK & F\_25-34\_US & F\_25-34\_!native\\
            \textbf{M\_25-34\_US} & M\_25-34\_US & F\_>45\_UK & M\_>45\_US & F\_25-34\_US & M\_25-34\_UK & M\_25-34\_!native \\
        \bottomrule
    \end{tabular}
    \caption{Summary of the condensed notation used to refer to listener treatment groups across speaker backgrounds, as discussed in Section~\ref{section:data-collection:collecting-perceived-sarcasm-labels}.}
    \label{table:condensed-notation}

\end{table}

We looked at the performance of each of the four groups in capturing sarcasm as intended by the speakers. We then compared the performance of each group to that achieved by the \emph{list=speak} group introduced in Section~\ref{section:data-collection:perceived-sarcasm:first-research-question}. For instance, for F\_25-34\_UK speakers, when the variable of investigation was \emph{age}, we looked at how well the perceived labels provided by F\_25-34\_UK listeners (same background as the speakers) matched the intended labels, compared to those provided by F\_>45\_UK listeners (age flipped). This allows us to quantify the influence of age in whether sarcasm is perceived as intended. Doing this for all variables of interest allows us to make a statement about our second research question, as discussed in Section~\ref{section:results-and-analysis:rq2}.
\subsubsection{Third Research Question}
\label{section:data-collection:perceived-sarcasm:third-research-question}
The third question asked if the influence of sociocultural variables on the effectiveness of sarcastic communication on OSNs is alleviated by the presence of contextual information. That is, for a given tweet, when the listeners have access to further cues to speaker intent, beyond the text of the tweet, do sociocultural variables still impact their ability to perceive the sarcastic nature of that tweet effectively, i.e. as intended by the speaker?

To investigate this, we first modified our label collection survey. The new version no longer showed tweet texts, but links to the corresponding tweets on Twitter. When labelling a tweet, we invited the listeners (survey participants) to consider not only the text of the tweet in the link, but also the surrounding tweets and any contextual information that they may find, either on the timeline, or on the profile of the speaker (the user who posted the tweet). 
We used two strategies to ensure that participants actually looked at contextual information found on Twitter. First, note that the modified survey only showed tweet links. They had to click the link, which would open a new tab in their web browser where they could see the text of the tweet on the corresponding Twitter page. Second, we manually checked the average response time per survey, which was around seven minutes longer for the modified survey, compared to the original survey.

We published the modified survey on Prolific Academic and collected perceived sarcasm labels for all tweets in our dataset from all six treatment groups mentioned thus far, with 3 labels per tweet from each group.
In a sentence where the specific background of a speaker is not relevant, we refer to listener treatment groups using a similar naming convention as above, while adding the prefix ``cont:'', as an abbreviation of ``context''. Then, the six groups are \emph{cont:list=speak} (listener group with the same background as the speaker), \emph{cont:list$\ne$speak} (listener group with background disjoint to that of the speaker), \emph{cont:list=speak-age}, \emph{cont:list=speak-gender}, \emph{cont:list=speak-country}, and \emph{cont:list=speak-native} (listener groups with the same background as the speaker except for flipping the variable of study, i.e. age, gender, country, or English language nativeness). Note that, for instance, \emph{list=speak} and \emph{cont:list=speak} refer to the same treatment group. The ``cont:'' prefix simply underlines that the group was asked to label tweets while also considering contextual information. As such, our group naming convention now includes an extra semantic layer, mainly a specification of the experimental setting in which labels were collected from that group: the absence of ``cont:'' indicates that the listeners were only shown tweet texts, while the inclusion of ``cont:'' indicates that they were shown tweet links and were asked to consider contextual information.

We compared intended labels with \emph{cont:list=speak} and \emph{cont:list$\ne$speak} perceived labels across our dataset, to see if listener sociocultural background still made a difference when contextual cues to speaker intent, found on speaker Twitter profiles, were considered by the listeners. Next, to study the potential influence of individual variables, we compared \emph{cont:list=speak} perceived labels to \emph{cont:list=speak-age}, \emph{cont:list=speak-gender}, \emph{cont:list=speak-country}, and \emph{cont:list=speak-native} perceived labels. This allows us to make a statement about our third research question, as discussed in Section~\ref{section:results-and-analysis:rq3}.

In summary, we have three research questions that we need to address. To address the first two questions, for each of the two speaker backgrounds, we collect labels from six treatment groups with 3 labels per tweet collected from each group to account for labelling noise. That amounts to 36 labels for each of the 120 tweets in our dataset. To address the third question, we collect 36 further labels per tweet from the same treatment groups, in a different experimental setting, where listeners are shown tweet links instead of tweet texts. That amounts to 72 labels for each tweet, giving a total of 8,640 labels collected.

This makes our analysis the first of its kind to sarcasm communication on OSN, with a sample size of at least an order of magnitude larger than any previous linguistic or psycholinguistic investigation.

We make the links to all tweets in our dataset, along with intended and perceived labels, publicly available for future analysis\footnote{\url{https://github.com/silviu-oprea/sarcasm-perception}}.
\subsection{Evaluation Metrics and Significance Testing}
\label{section:data-collection:evaluation-metrics-and-significance-testing}
As we saw, we have several treatment groups, each group including listeners of a specific sociocultural background. Answering our research questions requires us to perform two types of computations: (1) Given a treatment group, we need to quantify its performance in detecting sarcasm in our dataset as intended by the speakers. (2) Given two treatment groups, we need to quantify the difference in their performance and its statistical significance.

We discuss how we perform each of the two types of computations below.
\subsubsection{F-score to Quantify Performance}
Quantifying the performance of a group in detecting sarcasm as intended by the speakers reduces to checking the match between intended labels and the perceived labels that the group provides. Consider the following example.

Assume we have collected 4 sarcastic and 2 non-sarcastic tweets in our dataset. Let $l^{(i)}=[1, 0, 0, 1, 1, 1]^T$ be the vector of intended sarcasm labels for these tweets, each position corresponding to a tweet, where $0$ denotes the absence of sarcasm and $1$ the denotes its presence in that tweet. We then collect for these tweets the vector $l^{(p)}=[1, 0, 1, 0, 0, 0]^T$ of perceived sarcasm labels from a listener (annotator).

In this scenario, a straightforward measure of performance is accuracy. That is, the ratio of the number of correct perceived labels (i.e. perceived labels that are $1$ for tweets intended sarcastic and $0$ for those intended non-sarcastic) to the number of tweets in the dataset. In our example, that would be $2/6$, as the intended and perceived labels only match for the first two tweets. However, accuracy can be misleading in scenarios such as ours, when dealing with imbalanced data, i.e. data where not all classes have the same number of representatives. Recall that we have a ratio of 1:4 of sarcastic to non-sarcastic tweets in our dataset. To see this, say a listener carelessly labels all tweets they see as sarcastic. In such a scenario, the accuracy achieved by that listener in our example above would be $4/6$, giving us false confidence in the ability of that listener to recognise intended sarcasm.

To avoid such a scenario, we use f-score instead of accuracy as a measure of the match between intended and perceived labels. F-score is, to our knowledge, the most popular metric used to measure the performance of classification systems in machine learning and natural language processing literature, due to its robustness to imbalanced data.

We now describe how f-score is computed. We start by defining the precision of sarcasm detection as
$$
p=\frac{\sum_{n=1}^6 l^{(i)}_n l^{(p)}_n}{\sum_{n=1}^6 l^{(p)}}.
$$
That is, the ratio of the number of times the listener said the a tweet was sarcastic and was correct (i.e. the perceived label was the same as the intended label), divided by the number of times they said it was sarcastic. In our example above $p=1/2$. Next we define recall as
$$
r=\frac{\sum_{n=1}^6 l^{(i)}_n l^{(p)}_n}{\sum_{n=1}^6 l^{(i)}}.
$$
That is, out of the total number of tweets intended as sarcastic, how many the listener got right. In our example $r=1/4$.
Finally, we define f-score as:
$$
f=\frac{2pr}{p+r}.
$$
That is, the harmonic mean of precision and recall. Note that f-score penalises large differences between precision and recall, making it robust to imbalanced data. We invite the interested reader to consult~\citet{f-score-2009} for a more in-depth discussion. In our example, $f=0.333$.
The higher the $f$-score, the better the listener is at perceiving sarcasm as intended by the speaker.
\subsubsection{Randomization Test to Compare Performance}
As we just saw, for each treatment group, we can compute the f-score between intended sarcasm labels and the perceived labels provided by that group. We interpret that f-score as a numerical summary of the performance of the group in capturing intended sarcasm. As such, given two groups, we can compute the corresponding f-scores, and quantify the difference in performance between the two groups as the numerical difference between the f-scores. However, in our experiments, while we do control the sociocultural background of each treatment group, there are many other variables that we are unable to control, for instance the level of focus of each listener, or their honesty. One attempt to account for such sources of noise is the fact that we collect three separate labels from each group. However, this does not provide sufficient grounds to believe noise is no longer a threat. As such, we would like to make a rigorous statement about the significance of the difference in f-score, in a framework that deals with uncertainty by design. The standard way to accomplish such a task is to use a statistical test of significance. In this work we use a randomisation test~\citep{randomization-test-1989}. \citet{randomization-test-2000} provide implementation details, along with an excellent formal and experimental argument as to why this test is appropriate when the metric of investigation is f-score. We encourage the interested reader to consult their work. Following them, we use a p-value threshold of $0.05$, and our null hypothesis states that the difference is not significant.

For brevity, in the rest of the paper, we employ the convention of omitting references to the randomisation test when characterising the difference between the performance of two treatment groups as significant or not. As such, we will say ``the difference is significant'' to mean ``the difference is statistically significant under a randomisation test with $0.01< p\leq 0.05$''. Similarly, we will say ``the difference is very significant'' when  $p\leq0.01$.
%
%
\section{Results \& Analysis}
\label{section:results-and-analysis}
Our results are reported in two tables. Table~\ref{table:simple-results} shows, for each speaker, the precision, recall, and f-score achieved by the six treatment groups that we consider in the first experimental setting, when shown tweet texts for labelling. Table~\ref{table:context-results} shows, for each speaker, the f-score achieved by the same treatment groups in the second setting, when only shown tweet links and asked to consult contextual information on Twitter. As discussed in Section~\ref{section:data-collection:collecting-perceived-sarcasm-labels}, information from Table~\ref{table:simple-results} will help us address research questions 1 and 2, while that in Table~\ref{table:context-results} will be used to address question 3.
The first row in each table shows the results achieved by the group \emph{list=speak}, i.e. the group of listeners who have the same sociocultural background as the speakers. The next five rows show the results achieved by the other groups. In these five rows, the value of a metric (precision, recall, or f-score) could be shown with an ``*'' symbol appended. This indicates a significant difference between the value achieved by the corresponding treatment group and the value achieved by \emph{list=speak} for that metric (c.f. discussion on statistical significance, Section~\ref{section:data-collection:evaluation-metrics-and-significance-testing}). If two ``*'' symbols are appended, the difference is very significant. For instance, in Table~\ref{table:simple-results}, the second row shows the results for the treatment group \emph{list$\ne$speak}. When speaker background is F\_25-34\_UK, \emph{list$\ne$speak} denotes the group M\_>45\_US. We notice that this group achieves a precision of 0.455, which is very significantly different to that achieved by the group \emph{list=speak}, i.e. F\_25-34\_UK, of 0.648.

Below there are three subsections, one for each of our research questions. Each subsection discusses in detail those results that are relevant for addressing the corresponding research question.
\subsection{RQ 1: Does Sociocultural Background Similarity Have an Influence?}
\label{section:results-and-analysis:rq1}
To address this question, we consider the first two rows from Table~\ref{table:simple-results}, corresponding to treatment groups \emph{list=speak} and \emph{list$\ne$speak}.

Consider speaker background F\_25-34\_UK first. In this case, \emph{list=speak} denotes group F\_25-34\_UK of listeners, and \emph{list$\ne$speak} denotes group M\_>45\_US. We notice a very significant drop in precision from 0.648 to 0.455 between the first and the second group ($p=0.00005$), but a small, insignificant drop in recall from 0.633 to 0.622 ($p=0.499$). This amounts to a very significant drop in f-score from 0.640 to 0.526 ($0.002$). This suggests a very significant influence of the sociocultural variables of investigation on the ability of listeners to perceive sarcasm as intended by female speakers from the UK between 25- and 34-years-old. The very significant variation in precision, with insignificant variation in recall, could suggest a higher predisposition of M\_>45\_US listeners to classifying a tweet as sarcastic, compared to F\_25-34\_UK listeners.

Next, consider speaker background M\_25-34\_US. In this case,  \emph{list=speak} denotes group M\_25-34\_US of listeners, and \emph{list$\ne$speak} denotes group F\_>45\_UK. We notice a significant drop in precision from 0.460 to 0.356 (p=0.011) between the first and the second group, but an insignificant drop in recall from 0.511 to 0.467 ($p=0.307$). This amounts to a drop in f-score from 0.484 to 0.404 that is still insignificant.
Overall, there seems to be a significant influence of the sociocultural variables of investigation when precision is the metric of interest. Similarly to the previous paragraph, the lower precision and the insignificant variation in recall could suggest a higher predisposition of F\_>45\_UK listeners to classifying a tweet as sarcastic, compared to M\_25-34\_US listeners. Considering the information in both paragraphs, it seems that listeners over 45-years-old, irrespective of gender and country, show a higher predisposition to considering a tweet sarcastic.

Comparing the results across the two speaker backgrounds, for \emph{list=speak} listeners, we notice a further aspect. Mainly, F\_25-34\_UK listeners labelled tweets coming from F\_25-34\_UK speakers with a higher f-score of 0.640, compared to only 0.484 achieved by M\_25-34\_US listeners when labelling tweets coming from M\_25-34\_US speakers. Sarcastic communication seems more effective between UK females than between US males. UK females seem to be better at understanding each other's sarcasm.

To sum up, the sociocultural variables of interest seem to significantly impact the effectiveness of sarcastic communication. It particular, the effectiveness seems significantly higher when interlocutors share the same background. This provides significant statistical ground for positively answering our research question. Furthermore, as side effects of our experiment, we noticed a higher predisposition of older listeners to interpret a tweet as sarcastic, and a more effective sarcastic communication between UK females than between US males.
\subsection{RQ 2: Which Sociocultural Variables are Most Influential?}
\label{section:results-and-analysis:rq2}
To address this question, we consider the performance of each of the treatment groups \emph{list=speak-age}, \emph{list=speak-gender}, \emph{list=speak-country}, and \emph{list=speak-native}, found in the last four rows in Table~\ref{table:simple-results}, to that of the group \emph{list=speak}, found in the first row. We are interested in how the performance changes as we flip each of the variables of interest.

Consider speaker background F\_25-34\_UK first. In this case, \emph{list=speak} denotes treatment group F\_25-34\_UK, and \emph{list=speak-age} denotes group F\_>45\_UK. We notice a very significant drop in precision from 0.648 to 0.483 ($p=0.0005$) between the two groups, an equal recall of 0.633, amounting to a significant drop in f-score from 0.640 to 0.548 ($p=0.017$). Here, listener age seems to exert a significant influence on the effectiveness of sarcastic communication between the listeners and the speakers in our experiment. Looking at the next treatment groups, \emph{list=speak-gender} which denotes M\_25-34\_UK, and \emph{list=speak-country} which denotes F\_25-34\_US, we do not notice any significant difference. 
We find the lack of a significant effect of county particularly intriguing. It seems that US females are statistically just as able to recognise the sarcasm of UK females as other UK females are. UK females may be using a flavour of sarcasm that is more apparent to listeners of both nationalities.
English language nativeness, on the other hand, seems significantly influential. Looking at the last row, we notice a very significant drop in the precision achieved by F\_25-34\_UK listeners, compared to that achieved by F\_25-34\_!native listeners, from 0.648 to 0.491 ($p=0.0003$). The change in recall is insignificant, from 0.633 to 0.622.
The overall drop in f-score from 0.640 to 0.549 is very significant ($p=0.01$).

Next, consider speaker background M\_25-34\_US. In this case, \emph{list=speak} denotes treatment group M\_25-45\_US, and \emph{list=speak-age} denotes group F\_>45\_UK. Interestingly, for tweets posted by speakers of the current background, we do not notice any significant influence of listener age. Sarcastic communication seems to have, statistically, the same level of effectiveness between younger US males, as it does between younger and older US males. Gender, on the other hand, seems to have a significant influence when speaker background is M\_25-34\_US. Indeed, comparing the performance of \emph{list=speak} which here denotes M\_25-34\_US, to that of \emph{list=speak-gender}, which here denotes F\_25-34\_US, we notice no significant change in precision, but a significant increase in recall from 0.511 to 0.633 ($p=0.034$). Young US females seem to be better at pointing out the sarcasm of young US males than other young US males are. Country does not seem to have an influence. The next variable with a significant influence is English language nativeness. Indeed, we notice a very significant drop in precision between M\_25-34\_US and M\_25-34\_!native treatment groups, from 0.460 to 0.355 ($p=0.01$).

To sum up, age seems to very significantly impact the effectiveness of sarcastic communication among UK females, but not among US males. That is, among UK females, age determines a social partitioning, perhaps each partition being characterised by a specific flavour of sarcasm. This does not seem to be the case among US males. On the other hand, sarcastic communication between genders seems to be more efficient in the UK compared to the US.
Country seems to not be influential. English language nativeness, on the other hand, does have a significant impact, irrespective of the speaker background considered. 
Our results provide statistical grounds for answering our second research question in the following way. Age, gender, and English language nativeness of the interlocutors, may have a significant influence on the effectiveness of sarcastic communication online. Consulting the corresponding p-values, in our experiment, age was the most influential, followed by English language nativeness, and gender.
\subsection{RQ 3: Are Sociocultural Variables Influential When Context is Provided?}
\label{section:results-and-analysis:rq3}
To address this question, we consult Table~\ref{table:context-results}. We compare the performance achieved by the treatment groups \emph{cont:list$\ne$speak}, \emph{cont:list=speak-age}, \emph{cont:list=speak-gender}, \emph{cont:list=speak-country}, and \emph{cont:list=speak-native}, found in the last five rows, to that of the group \emph{cont:list=speak}, found in the first row. We are interested in whether there is any significant performance variation between \emph{cont:list=speak} and any of the other five treatment groups. If there is, this would indicate that sociocultural variables may still have an influence, even in the second experimental setting where listeners were only shown tweet links and were asked to consider contextual information found on Twitter.

Consider speaker background F\_25-34\_UK first. In this case, \emph{cont:list=speak} denotes treatment group F\_25-34\_UK, and \emph{cont:list$\ne$speak} denotes group M\_>45\_US. We notice a significant drop in precision between the two groups from 0.575 to 0.504 ($p=0.04$), with no significant changes in recall and f-score. The drop in precision is less, however, that it was in the first experimental setting, when listeners were shown tweet texts. The availability of contextual information seems to have alleviated, but not eliminated, the influence of listener sociocultural traits on their ability to recognise sarcasm as intended by the speakers. Let us consult the last four rows of Table~\ref{table:context-results} to see which traits remain influential. Comparing \emph{cont:list=speak-age}, which here denotes F\_>45\_UK, to \emph{cont:list=speak}, we notice a very significant drop in precision, from 0.575 to 0.471 ($p=0.005$), an insignificant drop in recall, amounting to a very significant drop in f-score from 0.640 to 0.540 ($p=0.003$). While the drop in precision is still less than it was in the first experimental setting, age remains a decisive factor. As in the first setting, gender and country are not significant. Unlike the first setting, however, the influence of English language nativeness of the listeners seems to have been eliminated by allowing listeners access to contextual information. Indeed, the change in precision, recall, and f-score, between \emph{cont:list=speak} and \emph{cont:list=speak-native} is no longer statistically significant in this experimental setting.

Next, consider speaker background M\_25-34\_US. In this case, we notice that the presence of contextual information has, statistically, eliminated the influence of sociocultural variables. Listener background does not seem to significantly influence the listener's ability to understand the sarcasm of M\_25-34\_US speakers when context is present. Granted, no listener does a particularly remarkable job, as the maximum f-score achieved by any group is less than 0.6. The important note is, however, that contextual information seems significantly more indicative of sarcasm produced by M\_25-34\_US speakers, than of that produced by F\_25-34\_UK speakers, as it is able to eliminate the influence of all sociocultural variables. Perhaps Twitter users from the United States disclose more public information on their profiles than users from the United Kingdom do.

To sum up, when context is available, age seems to very significantly impact the effectiveness of sarcastic communication among UK females, but not among US males. The impact of all the other sociocultural variables investigated seems to be eliminated by the presence of context. This is the answer that our experiment suggests to the third research question. 
\begin{table}[t]
    \centering
    \small
    {\color{black}
        \begin{tabular}{@{}l|cccc|cccc@{}}
            \toprule
                 & \multicolumn{4}{c|}{\textbf{speaker F\_25-34\_UK}} & \multicolumn{4}{c}{\textbf{speaker M\_25-34\_US}}\\
                & \textbf{listener} & \textbf{prec.} & \textbf{rec.} & \textbf{f} &  \textbf{listener} & \textbf{prec.}     & \textbf{rec.} & \textbf{f}\\\midrule
                list=speak         & F\_25-34\_UK & 0.648 & 0.633 & 0.640 & M\_25-34\_US & 0.460 & 0.511 & 0.484\\\midrule
                list$\ne$speak & M\_>45\_US & 0.455** & 0.622 & 0.526* & F\_>45\_UK & 0.356* & 0.467 & 0.404\\                
                list=speak-age     & F\_>45\_UK & 0.483** & 0.633 & 0.548* & M\_>45\_US & 0.477 & 0.578 & 0.523\\
                list=speak-gender  & M\_25-34\_UK & 0.610 & 0.678 & 0.642 & F\_25-34\_US & 0.483 & 0.633* & 0.548\\
                list=speak-country & F\_25-34\_US & 0.582 & 0.633 & 0.606 & M\_25-34\_UK & 0.422 & 0.544 & 0.476\\
                list=speak-native  & F\_25-34\_!native & 0.491** & 0.622 & 0.549** & M\_25-34\_!native & 0.355** & 0.544 & 0.430\\
            \bottomrule
        \end{tabular}}
    \caption{Experimental results addressing research questions 1 and 2. In the first column we show the name of each treatment group. Next, for each speaker background, we shown precision, recall, and f-score results achieved by each treatment group. `*' indicates a significant difference (p-value threshold of 0.05) between the value achieved by the corresponding treatment group and the one achieved by \emph{list=speak}. `**' indicates a very significant difference (p-value threshold of 0.01).
    }
    \label{table:simple-results}
\end{table}
\begin{table}[t]
    \centering
    \small
    {\color{black}
        \begin{tabular}{@{}l|cccc|cccc@{}}
            \toprule
                 & \multicolumn{4}{c|}{\textbf{speaker F\_25-34\_UK}} & \multicolumn{4}{c}{\textbf{speaker M\_25-34\_US}}\\
                & \textbf{listener} & \textbf{prec.} & \textbf{rec.} & \textbf{f} &  \textbf{listener} & \textbf{prec.} & \textbf{rec.} & \textbf{f}\\\midrule
                cont:list=speak     & F\_25-34\_UK & 0.575 & 0.722 & 0.640 & M\_25-34\_US & 0.431 & 0.589 & 0.498\\\midrule
                cont:list$\ne$speak & M\_>45\_US & 0.504* & 0.744 & 0.601 & F\_>45\_UK & 0.406 & 0.644 & 0.498\\                
                cont:list=speak-age & F\_>45\_UK & 0.471** & 0.633 & 0.540** & M\_>45\_US & 0.403 & 0.578 & 0.475\\
                cont:list=speak-gender & M\_25-34\_UK & 0.583 & 0.622 & 0.602 & F\_25-34\_US & 0.483 & 0.644 & 0.552\\
                cont:list=speak-country & F\_25-34\_US & 0.606 & 0.700 & 0.649 & M\_25-34\_UK & 0.451 & 0.567 & 0.502\\
                cont:list=speak-native & F\_25-34\_!native & 0.500 & 0.722 & 0.591 & M\_25-34\_!native & 0.408 & 0.667 & 0.506\\
            \bottomrule
        \end{tabular}
    }
    \caption{Experimental results addressing research question 3. `*' indicates a significant difference (p-value threshold of 0.05) between the value achieved by the corresponding treatment group and the one achieved by \emph{cont:list=speak}. `**' indicates a very significant difference (p-value threshold of 0.01).}
    \label{table:context-results}
\end{table}
\section{Discussion}
\label{section:discussion}
In this section we summarise the answers that Section~\ref{section:results-and-analysis} suggests to our research questions, discuss what implications these answers could have for future work, and conclude with what we believe to be key takeaways from this paper.
\subsection{Answers to Research Questions}
In Section~\ref{section:introduction} we introduced three research questions. To our knowledge, our work is the first to provide a quantitative investigation into these questions. Furthermore, we believe to also be the first to quantitatively investigate such questions through the lens of OSNs, often the deployment platform of social analysis tools designed within the CSCW community.

We first asked if the effectiveness of sarcastic communication is influenced by whether the interlocutors have similar sociocultural backgrounds. The investigation in Section~\ref{section:results-and-analysis:rq1} suggests a positive answer to this question. The sociocultural variables that we investigated, age, gender, country, and English language nativeness, had a statistically significant influence. As a side effect of that investigation, we noticed a more effective sarcastic communication between UK females than between US males. Furthermore, we argued that UK females may use a more apparent flavor of sarcasm, recognised better by all listeners.
One could view this as evidence in support of Utsumi's Implicit Display Theory~\citep{utsumi-2000} (c.f. Section~\ref{section:socio-cultural-variables:implcit-display-theory}) in that sarcasm is prototype-based category. That is, there is a concept of prototypical sarcasm which utterances can express to varying degrees. In other words, an utterance can be more or less sarcastic.

Our second question asked which sociocultural variables have the highest influence. The investigation in Section~\ref{section:results-and-analysis:rq2} suggests that the most influential variable is age, followed by English language nativeness, and gender. The significant influence of English language nativeness that we notice is in accordance with previous work that points out differences in the meanings native and non-native users of English attribute to conversational implicatures~\citep{bouton-1988,bouton-1992}.

Our final question was whether the presence of contextual information alleviates the influence of the variables discussed, in light of research such as~\citep{coolness} which conjectures that it might, but does not explore if that is actually the case. The investigation in Section~\ref{section:results-and-analysis:rq3} suggests that age remains influential on the effectiveness of sarcastic communication among UK females, but not among US males. The influence of all other sociocultural variables seems to be eliminated. We also noted that contextual information seems to be more indicative of the sarcasm produced by US males, than of that produced by UK females, perhaps suggesting that US males disclose more information on their Twitter profiles than UK females do.
\subsection{Key Takeaways}
Here we summarise what we believe to be the key takeaways from our results. We showed that the sociocultural background of the interlocutors may influence the effectiveness of their sarcastic exchanges on Twitter. This suggests that such background information should be considered in the design of future social analysis tools that either study sarcasm directly, or look at related phenomena where sarcasm may have an influence~\citep{sarcasm-in-sentiment-analysis}, such as the expression of sentiment, emotion, and hate-speech.

We provided a statistical methodology for comparing the significance of specific sociocultural variables. The most influential variables were English language nativeness, age, and gender, in this order. We also showed that public Twitter information can provide enough contextual cues to speaker intent to eliminate the influence of all sociocultural variables investigated except for age. Again, this suggests that such contextual cues should be considered in the design of future social investigations of sarcasm or related phenomena.

We made observations regarding the online social ecology surrounding sarcastic discourse. However, we believe future qualitative investigation that is out of the scope of this paper (i.e. not directly related to our research questions) is necessary to verify these observations. Mainly, we noted more effective sarcastic exchanges between UK females than between US males. We also noted that UK females may use a more apparent form of sarcasm than US males, that is easier to detect for listeners of both nationalities. Consistent with this, we observed contextual information to be more indicative of the sarcasm of US males. Our results also suggested a more effective sarcastic communication across genders in the UK compared to the US, but more effective across age groups in the US compared to the UK.

Finally, the fact that sarcasm used by UK females seemed easier to detect for listeners of both nationalities could be an argument in favour of Utsumi's theory of sarcasm \citep{utsumi-2000} (cf. Section \ref{section:socio-cultural-variables:implcit-display-theory}) in that there is a concept of prototypical sarcasm which utterances can express to varying degrees.
As in the previous paragraph, however, we believe these observations require further qualitative investigation that is out of the scope of this paper.
\subsection{Implications for Future Work}
We discuss two main ways in which we believe our work could inform future research, and suggest potential ways forward.
\subsubsection{Design of Social Analysis Tools}
As discussed in the previous section, our findings indicate that both the sociocultural variables investigated, and public social information, may be informative in design of social analysis tools that investigate sarcasm or related phenomena. We suggest a few ways in which all this information may be procured when exploring the Twitter network, given the popularity of tweet datasets. Public social information is easily accessible manually, or programatically, using the Twitter Application Programming Interface (API). The sociocultural variables are usually either available, or can be inferred from, public profile information. If inference is necessary, \citet{infer-age-from-twitter} suggest how to infer age of Twitter users based on whom they follow. \citet{infer-location-from-twitter} use a Bayes model coupled with a convolutional network to infer the location of timeline tweets. The country from which most timeline tweets originate may be considered the user's country. \citet{infer-gender-from-twitter} use a boosted stacked classifier to detect gender of Twitter users. English language nativeness could be deduced from the language of most timeline tweets, in conjunction with (available or inferred) user location. Once these variables are inferred, they can be either manually explored, or encoded in a computational framework. If encoding is required, one could identify a certain trait with the embedded representation of a set of tweets that come from users who posses that trait. For instance, the trait of being female could be encoded as the joint embedding of a set of tweets that all come from female users. The embedding could be built, for instance, using the ParagraphVector model~\citep{doc2vec,oprea-magdy-2019-exploring}.
\subsubsection{Usage of the Experimental Setup for Analysing Other Phenomena}
Our experimental setup could be used to study how the sociocultural traits of interlocutors influence the usage and interpretation of other linguistic phenomena, such as metaphors; or of social phenomena, such as hate speech and fake news. To this end, we provide as open science the web applications we developed that host our surveys for data collection and labelling, along with the software we wrote for data aggregation, reporting, and significance testing\footnote{Interested readers are welcome to contact us. We will be happy to provide all resources and assist with the setup.}.
%
%
\section{Conclusion \& Future Work}
\label{section:conclusion}
In this paper we have considered how sarcastic communication in OSNs can be influenced by the sociocultural backgrounds of the interlocutors. We asked whether similar backgrounds lead to more effective communication, which sociocultural variables have the most influence on the effectiveness, and whether the influence is alleviated by the presence of contextual information found publicly on the OSNs.

Consulting psycholinguistic studies of sarcastic communication, as well as linguistic theories as sarcasm, we chose four variables for investigation: gender, age, country, and English language nativeness.
For our experiments, we collected sarcastic tweets from Twitter users who posted them (whom we call \emph{speakers}), implicitly labelled by the users themselves (intended labels). We then had third-party annotators (whom we call \emph{listeners}) further label these tweets for sarcasm (perceived labels). Finally, we compared intended and perceived labels using f-score as a quantifier for similarity. 
Our results indicate that age, English language nativeness, and gender are statistically influential. The influence of age is maintained even when contextual information is available. We suggest that these variables, along with public social information, should be included in the future design of social analysis tools that either investigate sarcasm directly, or look at related phenomena where sarcasm may have an influence, such as the expression of sentiment, emotion, and hate-speech.
We also made observations regarding social behaviour. We noted a more effective sarcastic communication across genders in the UK compared to the US, but more effective across ages in the US compared to the UK. Furthermore, we noted that UK females may use a more apparent form of sarcasm, compared to the more subtle sarcasm of US speakers. Finally, contextual information seemed more indicative of the sarcasm of US males than of that of UK females.

In future work we plan to address the main limitations of the current work. First, despite being the largest study of its kind to out knowledge, we still only investigate two speaker backgrounds, F-25-34-UK and M-25-34-US. We plan to explore more in the future. Second, we plan to account for potential variations in the usage of sarcasm across the United States. Finally, we intend to study potential interactions between sociocultural variables. To make a statistically significant claim in this direction, we need labels from all possible sociocultural backgrounds spanned by the variables we consider, which we plan to collect in the future.
%
%
\section{Acknowledgements}
This work was supported in part by the EPSRC
Centre for Doctoral Training in Data Science,
funded by the UK Engineering and Physical Sciences Research Council (grant EP/L016427/1);
the University of Edinburgh; and The Financial
Times.
\bibliographystyle{ACM-Reference-Format}
\bibliography{main}
\end{document}